\documentclass[aps,prb,reprint,twocolumn,notitlepage,nofootinbib]{revtex4-1}

\usepackage{amsmath,amssymb,amsthm,graphicx,mathtools,siunitx,braket,bm}
\usepackage[varg]{txfonts}
\usepackage[version=4]{mhchem}

\allowdisplaybreaks

\begin{document}
	\title{Finite-Size Effects in the Dynamic Conductivity and Faraday Effect \\ of Quantum Anomalous Hall Insulators}
	\author{Junjie Zeng}
	\author{Tao Hou}
	\author{Zhenhua Qiao}
	\email[Corresponding author:~~]{qiao@ustc.edu.cn}
	\affiliation{ICQD, Hefei National Laboratory for Physical Sciences at Microscale, Synergetic Innovation Center of Quantum Information and Quantum Physics, CAS Key Laboratory of Strongly-Coupled Quantum Matter Physics, and Department of Physics, University of Science and Technology of China, Hefei, Anhui 230026, China}
	\author{Wang-Kong Tse}
	\email[Corresponding author:~~]{wktse@ua.edu}
	\affiliation{Department of Physics and Astronomy, The University of Alabama, Tuscaloosa, Alabama 35487, USA}
	\affiliation{Center for Materials for Information Technology, The University of Alabama, Alabama 35401, USA}
	
	\begin{abstract}
		We theoretically study the finite-size effects in the dynamical response of a quantum anomalous Hall insulator in the disk geometry. Semi-analytic and numerical results are obtained for the wavefunctions and energies of the disk within a continuum Dirac Hamiltonian description subject to a topological infinite mass boundary condition. Using the Kubo formula, we obtain the  frequency-dependent longitudinal and Hall conductivities and find that optical transitions between edge states contribute dominantly to the real part of the dynamic Hall conductivity for frequency values both within and beyond the bulk band gap. We also find that the topological infinite mass boundary condition changes the low-frequency Hall conductivity to $ e^2/h $ in a finite-size system from the well-known value $ e^2/2h $ in an extended system. The magneto-optical Faraday rotation is then studied as a function of frequency for the setup of a quantum anomalous Hall insulator mounted on a dielectric substrate, showing both finite-size effects of the disk and Fabry-P\'erot resonances due to the substrate. Our work demonstrates the important role played by the boundary condition in the topological properties of finite-size systems through its effects on the electronic wavefunctions.
	\end{abstract}
	
	\maketitle
	
	\section{Introduction}
	Topological properties are usually studied in extended systems without a confining boundary condition. As the example of integer quantum Hall effect illustrates, edge states existing in a realistic finite-size geometry are indispensable to the explanation of the underlying quantization phenomenon. In this connection, models with exactly solvable edge state wavefunctions are particularly valuable in delineating the role played by edge states in transport and optical phenomena \cite{Wunsch2008,Lu2010,Shan2010,Grujic2011,Christensen2014PRB}. The quantum anomalous Hall insulator \cite{Liu2016ARoCMP} is a two-dimensional topological state of matter characterized under electrical transport conditions by a quantized value of Hall conductivity and a zero longitudinal conductivity due to spin splitting under broken time-reversal symmetry. For the particular case when the quantized value is an integer multiple of the conductance quantum, the system has an integer Chern number and is also called a Chern insulator \cite{Haldane1988PRL}. A half-quantized Hall conductivity is also possible \cite{Qi_PRB2008}, as is the case for the surface states of a three-dimensional topological insulator. Quantum anomalous Hall insulator has been experimentally realized in magnetically doped three-dimensional topological insulator thin films \cite{Chang2013S}. Two-dimensional atomically thin materials doped with heavy magnetic adatoms \cite{Ren2016RoPiP} have also been proposed as platforms for realizing quantum anomalous Hall insulator \cite{Qiao2010PRB,Tse2011PRBa,Deng2017PRB}. This class of system has the advantage that they provide an additional tunability of topological phases due to interplay between the magnetization and an applied out-of-plane electric field \cite{Qiao_Tse_PRL2011,Qiao_Tse_PRB2013,Zeng2017PRB}.
	
	Frequency-dependent conductivity provides a useful probe for charge carriers' dynamical response and elementary excitations. In systems where the electron's momentum is coupled to the pseudospin or spin degrees of freedom, it reveals unusual interaction renormalization \cite{Tse_MD_PRB2009,Li_Tse_PRB2017} and strong-field \cite{Lee_Tse_PRB2017} effects. The complex dynamic Hall conductivity can yield valuable information in topological materials beyond the direct current limit, with its real part providing the dynamics of the reactive carrier response and imaginary part the dissipative response. Dynamic Hall response can be probed optically by the magneto-optical Faraday and Kerr rotations. Three-dimensional topological insulator thin films under broken time-reversal symmetry exhibit dramatic Faraday and Kerr effects in the low-frequency regime that signifies the underlying topological quantization of the Hall conductivity \cite{Tse2010PRL,Maciejko_PRL2010,wu2016quantized,okada2016terahertz,dziom2017observation}. The dynamic Hall response and magneto-optical effects of topological materials are often theoretically studied assuming an infinitely extended system, and so far there has been few study on the finite-size effects of these properties due to the finite planar dimensions of the system. It is the purpose of this work to perform a semi-analytical and numerical study of the dynamic conductivities and magneto-optical Faraday effect of quantum anomalous Hall insulator in a finite circular disk geometry.
	
	Our theory is based on the low-energy continuum description of quantum anomalous Hall insulator subject to a vanishing radial current boundary condition (`no-spill' boundary condition). The system is described by a massive Dirac Hamiltonian, which would give a half-quantized Hall conductivity when the system is infinitely extended. We compute the exact energies and wavefunctions of the finite disk as a function of the orbital angular momentum quantum number, and use the Kubo formula to evaluate the dynamic longitudinal and Hall conductivities. Contrary to the extended system case, we find that imposing a change of the topological character of the system across the boundary through the no-spill boundary condition causes the finite-sized massive Dirac model to carry a Chern number of $ 1 $ instead of $ 1/2 $. Our calculations also show that optical dipole transitions between edge states contributes to an almost constant value of $ e^2/h $ in the dynamic Hall conductivity that remains constant even for frequencies exceeding the band gap. Combining all three types of optical transitions among the edge and bulk states, the total dynamic conductivities of the finite disk are found to agree with the main features calculated from the massive Dirac model with an additional parabolic dispersion term that breaks electron-hole symmetry in an extended system. The finite-size effects in the conductivities are also seen in the magneto-optical Faraday rotations. Here, we also study the `finite-size effects' along the out-of-plane direction by considering a substrate interfacing the quantum anomalous Hall insulator. The finite thickness of the substrate gives rise to Fabry-P\'erot oscillations of the Faraday effect, which are found to exert a stronger influence on the Faraday rotation spectrum than the effect of the finite disk size.
	
	This article is organized as follows. In Sec.~\ref{sec:model} we describe the model Hamiltonian and boundary condition of the quantum anomalous Hall insulator disk and derive semi-analytic expressions for the eigenfunctions and energies. Sec.~\ref{sec:conduc} describes our calculations and results for the dynamic longitudinal and Hall conductivities using the Kubo formula. A flow diagram plotting the real parts of the longitudinal and Hall conductivities is discussed in Sec.~\ref{sec:flow}, which approaches the behavior of a Chern insulator with increasing disk radius. In Sec.~\ref{sec:F}, we provide results for the magneto-optical Faraday rotation of the quantum anomalous Hall insulator disk and study the effects of finite disk radius and the presence of an underlying substrate. Sec.~\ref{sec:disc_conc} summarizes our work.
	
	\section{Theoretical Model}\label{sec:model}
\begin{figure}[htp!]
		\centering
		\includegraphics[width=.45\textwidth]{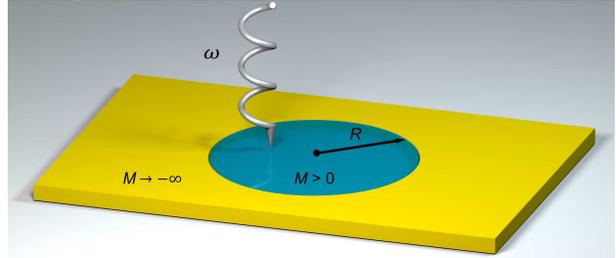}
		\caption{\label{fig:schematic} (Color online) Schematic of the system under investigation. The cyan (dark) area is a quantum anomalous Hall insulator nanodisk with a radius $ R $ and a positive finite mass in the Dirac model Eq.~(\ref{eq:Hamiltonian}), surrounded by a yellow (light) domain with a mass going to negative infinity, to ensure a topologically nontrivial domain wall. The outer region is infinitely large. The linear response dynamics of this system is studied in Sec.~\ref{sec:conduc} by initializing a weak perpendicular incident light with energy $ \omega $.}
	\end{figure}

	As depicted in Fig.~\ref{fig:schematic}, we consider a quantum anomalous Hall insulator in a disk geometry with a radius $ R $. The low-energy physics of a quantum anomalous Hall insulator is described by the massive Dirac Hamiltonian, 
	\begin{align}\label{eq:Hamiltonian}
		h_\text{D} &= v\ (\bm{\sigma}\times\bm{p})\cdot\hat{\bm{z}} + M\sigma_z,
	\end{align}
	where $ v $ is the band velocity of the Dirac model, $ M > 0 $ is the Zeeman interaction that breaks time-reversal symmetry and provides a bulk band gap, and $ \bm{\sigma}=(\sigma_x, \sigma_y, \sigma_z) $ is the 3-vector formed by the Pauli matrices corresponding to the spin degree of freedom. Under periodic boundary  conditions, the massive Dirac Hamiltonian gives a Chern number $ 1/2 $. To incorporate finite-size effects, we need to specify the appropriate boundary condition at the edge of the disk. While a vanishing wavefunction boundary condition is suitable for graphene with zigzag edge termination, there is no reason to assume that such a condition also applies in our case without appealing to atomistic details at the boundary. In particular, since we aim to provide a treatment of the finite-size effects of quantum anomalous Hall insulator as generally as possible, we use the no-spill current condition at the boundary, which is more suitable to be used with a continuum Hamiltonian model. We will discuss this boundary condition in more details.
	
	The wavefunctions and energy eigenvalues are obtained by solving the eigenvalue problem $ h_\text{D}\psi=\epsilon\psi $, where $ \psi $ is a two-component spinor. In view of the rotational symmetry of the system, the eigenvalue problem can be facilitated by using the following wavefunction ansatz in polar coordinates \cite{Christensen2014PRB}, 
	\begin{align}\label{eq:ansatz}
		\psi(\rho,\phi)&=
		\begin{pmatrix}
			\psi_\uparrow\\
			\psi_\downarrow
		\end{pmatrix}
		=\mathrm{e}^{\mathrm{i} l\phi}
		\begin{pmatrix*}[r]
			A_l(\rho)\\
			\mathrm{e}^{\mathrm{i}\phi}B_l(\rho)
		\end{pmatrix*},
	\end{align}
	where $ \rho $ is the radial position, $ \phi $ the azimuthal angle, and $ l $ is an integer corresponding to the orbital angular momentum quantum number.
	Combining Eq.~(\ref{eq:Hamiltonian}) and Eq.~(\ref{eq:ansatz}), one obtains the wavefunctions for states lying both outside and inside of the bulk energy gap
	\begin{align}\label{eq:wave_function}
		\begin{dcases}
			\Psi_{ln}(\rho,\phi)=\dfrac{\mathrm{e}^{\mathrm{i} l\phi}}{\sqrt{N_{ln}}}
			\begin{pmatrix*}[r]
				(\epsilon+M)\ \mathrm{J}_{l}(\beta_{ln}\tilde\rho)\\
				+\mathrm{e}^{\mathrm{i}\phi}\sqrt{\epsilon^2-M^2}\ \mathrm{J}_{l+1}(\beta_{ln}\tilde\rho)
			\end{pmatrix*}, & |\epsilon| > M\\
			\\
			\Phi_{ln}(\rho,\phi)=\dfrac{\mathrm{e}^{\mathrm{i} l\phi}}{\sqrt{\mathcal{N}_{ln}}}
			\begin{pmatrix*}[r]
				(M+\epsilon)\ \mathrm{I}_{l}(b_{ln}\tilde\rho)\\
				-\mathrm{e}^{\mathrm{i}\phi}\sqrt{M^2-\epsilon^2}\ \mathrm{I}_{l+1}(b_{ln}\tilde\rho)
			\end{pmatrix*}, & |\epsilon| < M
		\end{dcases}
	\end{align}
	where we have defined the system's characteristic frequency $ \omega_R\equiv v/R $, the dimensionless wavenumbers $ \beta=\sqrt{\epsilon^2-M^2}/(\hbar\omega_R) $ and $ b=\sqrt{M^2-\epsilon^2}/(\hbar\omega_R) $, as well as the scaled dimensionless radial distance $ \tilde{\rho}\equiv\rho/R\in[0,1] $. Here $ \text{J}_l(x) $ is the Bessel function and $ \text{I}_l(x) $ the modified Bessel function, both of the first kind. The normalization coefficients $ N_{ln} $ and $ \mathcal{N}_{ln} $ follow from the normalization condition of the wavefunctions $ 1=\int_{0}^{2\pi}\mathrm{d}\phi\int_{0}^{R}\mathrm{d}\rho\rho|\psi(\rho,\phi)|^2=2\pi R^2\int_{0}^{1}\mathrm{d}\tilde{\rho}\tilde{\rho}|\psi(\rho,\phi)|^2 $, and are consequently given by
	\begin{widetext}
		\begin{align}\label{eq:normalization_coeff}
			\begin{dcases}
				N_{ln}=2\pi R^2\epsilon_{ln}(\epsilon_{ln}+M)\left[\mathrm{J}_l^2(\beta_{ln})+\mathrm{J}_{l+1}^2(\beta_{ln})-\dfrac{2l+1-M/\epsilon_{ln}}{\beta_{ln}}\mathrm{J}_l(\beta_{ln})\mathrm{J}_{l+1}(\beta_{ln})\right], & |\epsilon| > M,\\
				\mathcal{N}_{ln}=2\pi R^2(M+\epsilon_{ln})\epsilon_{ln}\left[\mathrm{I}_l^2(b_{ln})-\mathrm{I}_{l+1}^2(b_{ln})-\dfrac{2l+1-M/\epsilon_{ln}}{b_{ln}}\mathrm{I}_l(b_{ln})\mathrm{I}_{l+1}(b_{ln})\right], & |\epsilon| < M.
			\end{dcases}
		\end{align}
	\end{widetext}
	
\begin{figure}[htp!]
		\centering
		\includegraphics[width=.47\textwidth]{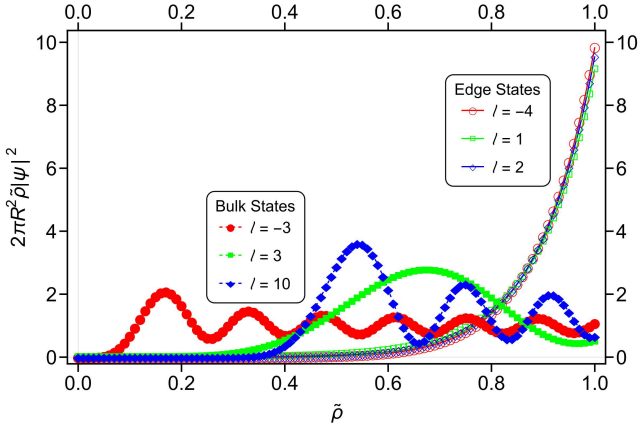}
		\caption{\label{fig:edge&bulk} (Color online) Radial probability density as a function of the scaled dimensionless radius $ \tilde{\rho}$ (with $ R=\SI{30}{\nano\meter} $) for several typical values of angular momentum $ l $ and energies $ \epsilon_{ln} $. Hollow-shape-dotted curves indicate the edge states $ \Phi_{ln} $ and solid-shape-dotted curves the bulk states $ \Psi_{ln} $.}
	\end{figure}

	In a finite-sized geometry, a clear delineation between bulk and edge states does not exist because there is always some degree of mixing between bulk and edge states. However, for the sake of terminology, we shall label $ \Psi_{ln} $ (having energies beyond the bulk band gap) and $ \Phi_{ln} $  (having energies within the bulk band gap) as the bulk and edge states, respectively. This terminology is supported by examining the radial probability density profile as a function of the scaled dimensionless radius $ \tilde{\rho} $ in Fig.~\ref{fig:edge&bulk}. It is seen that the in-gap states $ \Phi_{ln} $ displays a monotonically increasing profile towards the edge of the disk and indeed behaves like edge states. On the other hand, the out-of-gap states $ \Psi_{ln} $ exhibits oscillations as a function of the radial position $ \tilde{\rho} $ and can therefore be associated with bulk states.
	
	To determine the energy eigenvalues, we impose the no-spill boundary condition that the outgoing radial component of the current vanishes at the edge of the disk. Within the massive Dirac model, this condition corresponds to taking the mass term $ M $ outside the disk region to infinity and is sometimes called the infinite mass boundary condition \cite{Christensen2014PRB}. We note that in principle there can be two choices for the sign of the mass term, with $ M(\rho>R)\to\pm\infty $. Since we assume $ M(\rho\leq R) > 0 $, the choice $ M(\rho >R) \to -\infty $ is appropriate here for a topologically nontrivial domain wall, ensuring a change in the topological character across the disk boundary \cite{Trushin2016PRB,FZhang_PRB2012}. This results in the following constraint between the two components of the wavefunction \cite{Berry1987PotRSoLA,Peres2009JoPCM} $ \psi_\downarrow/\psi_\uparrow=\alpha\mathrm{e}^{\mathrm{i}\phi} $, with $ \alpha=-1 $, (cf. Appendix \ref{appsec:IMBC} for more information). This choice of $ M( \rho > R ) $ differs from that in the original Berry's paper \cite{Berry1987PotRSoLA}, which addressed a situation without a topological domain wall. For this reason, we refer to the boundary condition we use here as the \textit{topological infinite mass boundary condition}. Armed with the above, one has the following equations for the eigenenergies $ \epsilon_{ln} $ from Eq.~(\ref{eq:wave_function})
	\begin{align}\label{eq:massive_IM_spec}
		\begin{dcases}
			(\epsilon_{ln}+M)\,\mathrm{J}_{l}(\beta_{ln})=-\sqrt{\epsilon_{ln}^2-M^2}\,\mathrm{J}_{l+1}(\beta_{ln}), & |\epsilon| > M,\\
			(M+\epsilon_{ln})\,\mathrm{I}_{l}(b_{ln})=+\sqrt{M^2-\epsilon_{ln}^2}\,\mathrm{I}_{l+1}(b_{ln}), & |\epsilon| < M.
		\end{dcases}
	\end{align}
	Equation~(\ref{eq:massive_IM_spec}) is transcendental equations with multiple roots $ \epsilon_{ln} $ for each given $ l $, where $ n = 1, 2, \dots $ indicates the multiplicity. While a number of candidates for Chern insulators have been proposed in the literature  \cite{Deng2019a,Jin2018PRB,He2018ARoCMP,Liu2016ARoCMP,Garrity2013PRL,Yu2010S}, we choose the band parameters  corresponding to the quantum anomalous Hall insulator of a \ce{Bi2Se3} thin film with $v = \SI{5.e5}{\meter\cdot\second^{-1}}$ and a Zeeman energy $ 2M = \SI{0.1}{\electronvolt} $. As a low-energy effective theory, the massive Dirac model is valid up to a certain energy cutoff $ \epsilon_{\mathrm{c}} $. For concreteness we use  $ \epsilon_{\mathrm{c}} =  \SI{.3}{\electronvolt} $ as the energy cutoff, noting that the main findings  of our work are not dependent on the precise value of $ \epsilon_{\mathrm{c}}$ with $ \epsilon_{\mathrm{c}} \ll M $.  We therefore only seek the numerical roots of Eq.~(\ref{eq:massive_IM_spec}) within the range $ [-\epsilon_{\mathrm{c}}, \epsilon_{\mathrm{c}}] $.
	\begin{figure}[htp!]
		\centering
		\includegraphics[width=.47\textwidth]{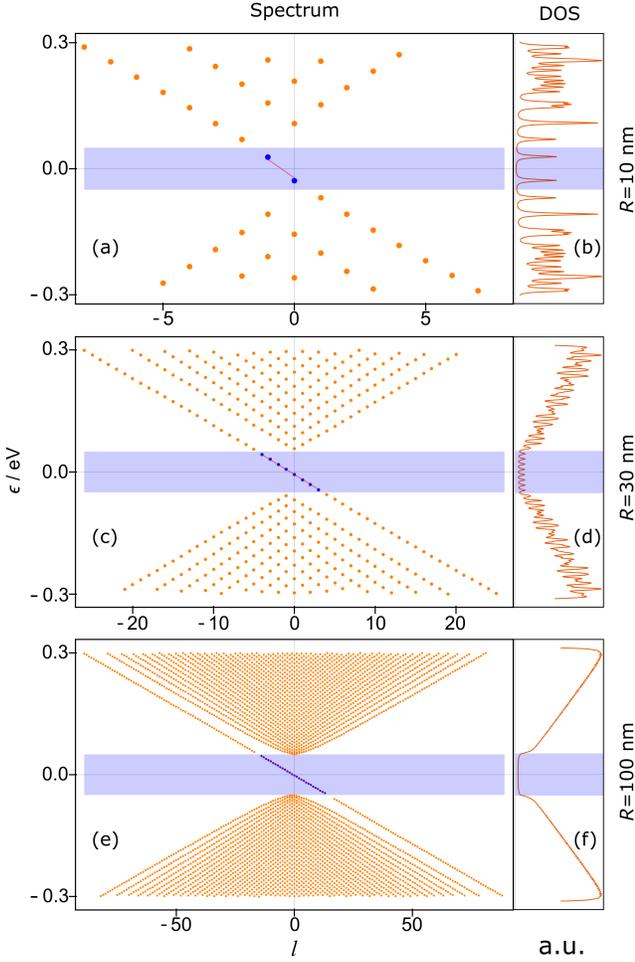}
		\caption{\label{fig:massive_spec&dos}(Color online) Energy spectrum (energy $ \epsilon $ versus angular momentum quantum number $ l $) and the corresponding density of states (DOS) for $ v=\SI{5.e5}{\meter\cdot\second^{-1}} $ and $ M=\SI{.05}{\electronvolt} $. The region shaded in light blue (gray) indicates the bulk energy gap, within which the red (darker) line shows the analytic dispersion Eq.~(\ref{eq:linear_disp}) of the chiral edge state.}
	\end{figure}
	
	Figures \ref{fig:massive_spec&dos} (a), (c) and (e) show the obtained energy spectrum for $ \epsilon $ as a function of $ l $ for different values of $ R $.
	One can identify features that correspond predominantly to bulk states and edge states. The bulk state spectrum contains a gap in which a one-way chiral edge dispersion runs across. As $ R $ increases from \SIrange{10}{100}{\nano\meter}, the delineation between the bulk and edge spectra becomes more evident. In the limit $ R\to\infty $ that can be achieved physically when $ R \gg \hbar v/M $, the chiral edge state dispersion can be obtained by expanding the second equation in Eq.~(\ref{eq:massive_IM_spec}) using a large $ R $ expansion:
	\begin{align}\label{eq:linear_disp}
		\epsilon_l &\simeq -\epsilon_0 \left(l+1/2\right),
	\end{align}
	with $ \epsilon_0 \equiv \hbar\omega_R[1+\hbar\omega_R/(2M)] $. We note that the $ 1/2 $ on the right hand side originates from the spin angular momentum of the electron. Gratifyingly, this approximate analytic dispersion is in excellent agreement with the exact numerical results as shown by  the red solid line in Fig.~\ref{fig:massive_spec&dos}. Using Eq.~(\ref{eq:linear_disp}), the number of the in-gap states can be estimated as $ \lfloor2M/(\hbar\omega_R)\rfloor $, where $ \lfloor\dots\rfloor $ denotes the floor function. Figures \ref{fig:massive_spec&dos} (b), (d) and (f) show the density of states of the calculated spectrum from the expression \cite{Datta2005}
	\begin{align}\label{eq:dos}
		D(\epsilon)&=\dfrac{1}{\pi\mathcal{A}}\sum_{\nu}\mathrm{Im}\dfrac{1}{\epsilon_\nu-\epsilon-\mathrm{i}\eta},
	\end{align}
	where `Im' stands for imaginary part, $ \mathcal{A} $ is the area of the disk and the broadening parameter is set as $ \eta=\SI{2.4e-3}{\electronvolt} $ throughout this work. The nonvanishing peaks of the density of states in the gap indicate the existence of the in-gap chiral edge states. When the radius increases, the DOS profile expectedly becomes smoother as more states are introduced into the system. On the other hand, a small radius ($ R \lesssim \hbar v/M $) enhances the quantum confinement effect as seen from the more prominent resonances from the individual quantum states.
	
	\section{Dynamic Conductivity}\label{sec:conduc}
	We now introduce into the system a weak, linearly polarized alternating current probe field that is normally  incident on the quantum anomalous Hall insulator disk. With the obtained energy spectrum and wavefunctions, we proceed to calculate the longitudinal and Hall optical conductivities using the Kubo formula \cite{Mahan2000} in the real space representation (derivation is provided in Appendix \ref{appsec:Kubo}):
	\begin{align}\label{eq:opt_cond_final}
		\sigma_{ij}(\omega)&=2\mathrm{i}\omega\frac{e^2}{h}\sum_{mm'}\dfrac{f_0'-f_0}{\Delta\epsilon-\omega-\mathrm{i}\eta}\dfrac{\braket{m|x_i|m'}\braket{m'|x_j|m}}{R^2},
	\end{align}
	where $ i,j \in \set{x,y} $, $ m (m') $ is a collective label for the relevant quantum numbers, $ f_0 $ is the Fermi-Dirac distribution function, and $ \Delta\epsilon=\epsilon'-\epsilon $
	is the energy difference between the final (primed) and the initial (unprimed) states in a transition, and $ \omega $ is the photon energy of the incident light.
	
	The matrix elements in Eq.~(\ref{eq:opt_cond_final}) capture the transition processes among the bulk states $ \Psi_{ln} $ and edge states $ \Phi_{ln} $ and there are three types of transitions, \textit{i.e.}, edge-to-edge (E-E), edge-to-bulk (B-E), and bulk-to-bulk (B-B). Using the expressions of the wavefunctions Eq.~(\ref{eq:wave_function}) together with their normalization coefficients Eq.~(\ref{eq:normalization_coeff}), we obtain the following matrix elements for the three types of transitions
	\begin{align}\label{eq:matrix_element}
		\Braket{\psi_{l'n'}^{S'}|\begin{pmatrix}
			x\\
			y
			\end{pmatrix}|\psi_{ln}^{S}}
		&=R\ \mathcal{I}_{l'n',ln}^{S'S}
		\begin{pmatrix*}[r]
			1 & 1\\
			\mathrm{i} & -\mathrm{i}
		\end{pmatrix*}
		\begin{pmatrix}
			\delta_{l,l'+1}\\
			\delta_{l,l'-1}
		\end{pmatrix},
	\end{align}
	where $ S', S\in\set{\text{B}, \text{E}} $ stand for bulk and edge states, $ \delta_{l,l'} $ is the Kronecker delta symbol, and $ \mathcal{I}_{l'n',ln}^{S'S} $ is a dimensionless radial integral defined by
	\begin{widetext}
		\begin{align}\label{eq:matrix_element_integrals}
			\mathcal{I}_{l'n',ln}^{S'S}=\int_{0}^{1}\mathrm{d}\tilde{\rho}\tilde{\rho}^2
			\begin{dcases}
				\dfrac{(M+\epsilon')(M+\epsilon)\ \mathrm{I}_{l'}(b_{l'n'}\tilde{\rho})\ \mathrm{I}_{l}(b_{ln}\tilde{\rho}) + \sqrt{(M^2-{\epsilon'}^2)(M^2-\epsilon^2)}\ \mathrm{I}_{l'+1}(b_{l'n'}\tilde{\rho})\ \mathrm{I}_{l+1}(b_{ln}\tilde{\rho})}{\sqrt{\mathcal{N}_{l'n'}\mathcal{N}_{ln}}/(\pi R^2)}, & S'=S=\text{E},\\
				\dfrac{(\epsilon'+M)(M+\epsilon)\ \mathrm{J}_{l'}(\beta_{l'n'}\tilde{\rho})\ \mathrm{I}_{l}(b_{ln}\tilde{\rho}) - \sqrt{({\epsilon'}^2-M^2)(M^2-\epsilon^2)}\ \mathrm{J}_{l'+1}(\beta_{l'n'}\tilde{\rho})\ \mathrm{I}_{l+1}(b_{ln}\tilde{\rho})}{\sqrt{N_{l'n'}\mathcal{N}_{ln}}/(\pi R^2)}, & S'=\text{B}, S=\text{E},\\
				\dfrac{(\epsilon'+M)(\epsilon+M)\ \mathrm{J}_{l'}(\beta_{l'n'}\tilde{\rho})\ \mathrm{J}_{l}(\beta_{ln}\tilde{\rho}) + \sqrt{({\epsilon'}^2-M^2)(\epsilon^2-M^2)}\ \mathrm{J}_{l'+1}(\beta_{l'n'}\tilde{\rho})\ \mathrm{J}_{l+1}(\beta_{ln}\tilde{\rho})}{\sqrt{N_{l'n'} N_{ln}}/(\pi R^2)}, & S'=S=\text{B}.
			\end{dcases}
		\end{align}
	\end{widetext}
	Equation~(\ref{eq:matrix_element}) expresses an angular momentum \emph{selection rule}: transitions are allowed only between states with a change in angular momenta $ \Delta l=\pm 1 $. The remaining radial integrations ($ \mathcal{I}_{l'n',ln}^{S'S} $) in Eq.~(\ref{eq:matrix_element_integrals}) are computed numerically.
	
\begin{figure*}[htp!]
		\centering
		\includegraphics[width=.85\textwidth]{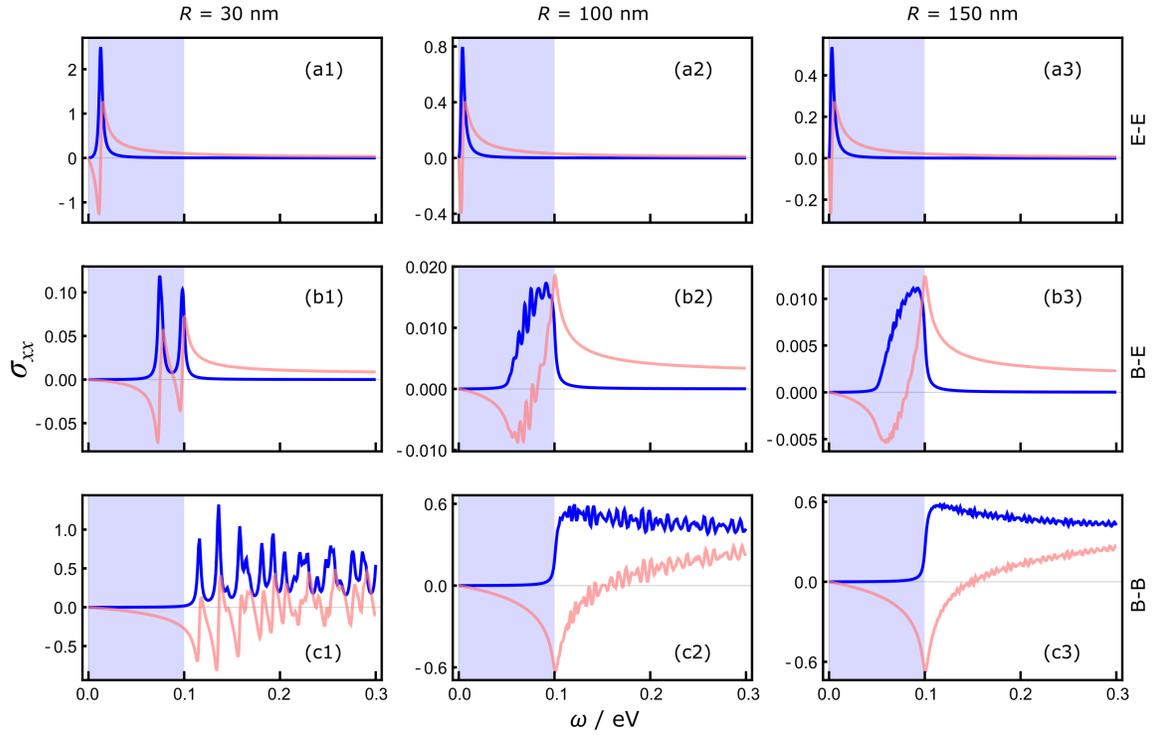}
		\caption{\label{fig:sigma_xx} (Color online) Plots of the longitudinal conductivity $ \sigma_{xx} $  (in unit of $ G_0 = e^2/h $) versus frequency $ \omega $ for different radius $ R $ and for contributions arising from different types of transitions. Blue (dark) indicates the real part and red (light) the imaginary part. The light blue (gray) regions indicate the extent of the bulk energy gap. Panels (a1)-(a3) depict the contribution from edge-to-edge (E-E) transitions, (b1)-(b3) the contribution from edge-to-bulk (B-E) transitions, and (c1)-(c3) the contribution from bulk-to-bulk (B-B) transitions. Values of $ v $ and $ M $ are the same as in Fig.~\ref{fig:massive_spec&dos}.}
	\end{figure*}

	Figure \ref{fig:sigma_xx} shows our results for the real (blue) and imaginary (red) parts of the longitudinal conductivity $ \sigma_{xx}(\omega) $ when the Fermi level $ \epsilon_\text{F} = 0 $ for different values of $ R $ separated into the three contributions: E-E (first row), B-E (second row) and B-B (third row). First, we note that the finite size of the disk has a different effect on the E-E conductivity contribution compared to the other two types of contributions involving the bulk.  As $ R $ is increased, the E-E conductivity [Figs.~\ref{fig:sigma_xx} (a1)-(a3)] remains approximately the same while the B-E and B-B contributions [Figs.~\ref{fig:sigma_xx} (b1)-(b3) and (c1)-(c3)] display considerable changes. According to the selection rule in Eq.~(\ref{eq:matrix_element}), there can only be one E-E transition along the chiral edge state dispersion from below to above the Fermi level, therefore there is always only one peak in the E-E conductivity regardless of the size of the disk. The peak's position is seen to shift toward $ \omega = 0 $ with increasing $ R $, because the edge states become more closely spaced with their energy separation $ \approx \epsilon_0 \propto 1/R $ [Eq.~(\ref{eq:linear_disp}), to first order]. In contrast, for the B-E [Figs.~\ref{fig:sigma_xx} (b1)-(b3)] and B-B contributions [Figs.~\ref{fig:sigma_xx} (c1)-(c3)], since the number of possible transitions is directly proportional to the number of bulk states, the number of peaks increases and the conductivity approaches a smooth continuous curve as $ R $ increases. The threshold beyond which the B-B contribution becomes finite corresponds to the bulk energy gap $ 2M $.
	\begin{figure*}[htp!]
		\centering
		\includegraphics[width=.87\textwidth]{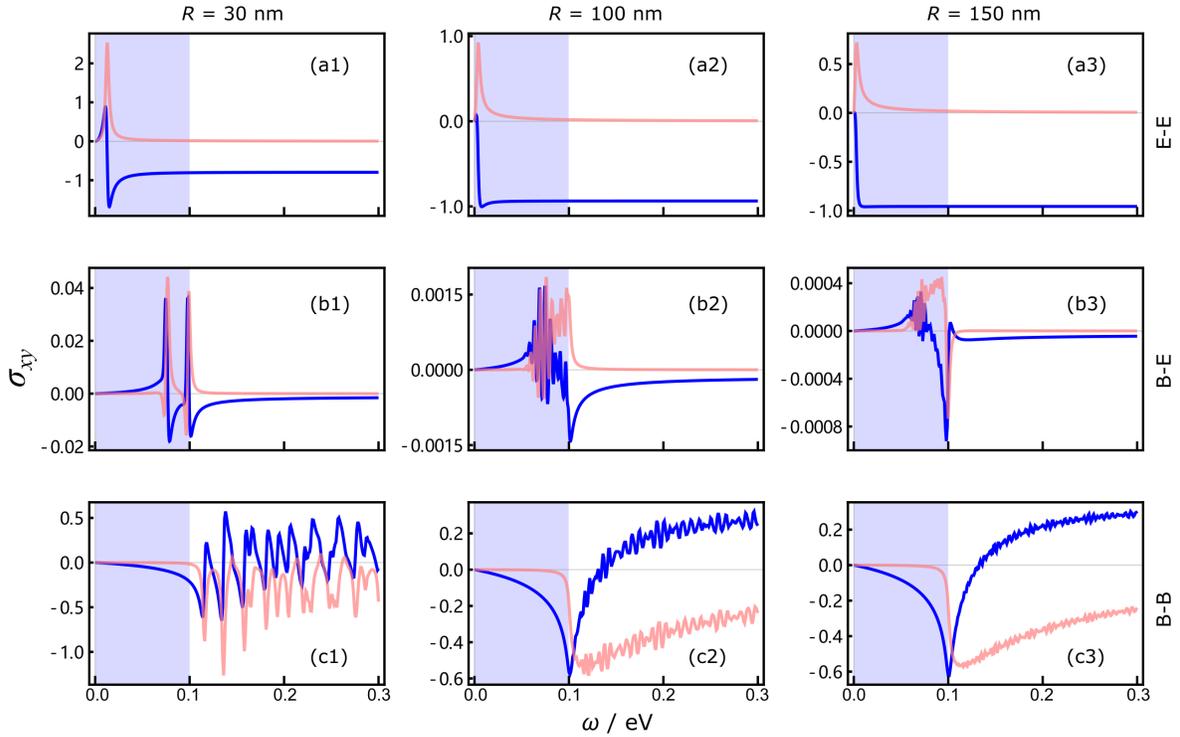}
		\caption{\label{fig:sigma_xy}(Color online) Hall conductivity $ \sigma_{xy} $ (in unit of $ G_0 = e^2/h $) versus frequency $ \omega $ for different radius $ R $ and for contributions arising from different types of transitions. Blue (dark) indicates the real part and red (light) the imaginary part.}
	\end{figure*}
	
	The corresponding dynamic Hall conductivity $ \sigma_{xy}(\omega) $ is shown in Fig.~\ref{fig:sigma_xy}. The above description for the longitudinal conductivity is also applicable here, if we note that the roles of the reactive and dissipative components are played by the real and imaginary parts of $ \sigma_{xy}(\omega) $ respectively.
	For both the longitudinal and Hall conductivities, the E-E contribution displays a smooth profile across all values of frequency, which is the result of only one possible edge-to-edge transition. The B-E contribution is smaller than both E-E and B-B contributions by an order of magnitude for the smallest radius $R = 30\,\mathrm{nm}$ and is further suppressed with increasing radius. For frequency within the bulk gap $2M$, its profile exhibits many closely spaced sharp peaks corresponding to the many possible edge-to-bulk transitions, and is smooth for frequency beyond the gap. The opposite behavior is seen in the B-B contribution. Its profile exhibits wild fluctuations due to even more possible bulk-to-bulk transitions for frequency beyond the bulk gap, which are suppressed when the disk radius is increased. There is an important distinction between Fig.~\ref{fig:sigma_xx} and Fig.~\ref{fig:sigma_xy}. As shown in Fig.~\ref{fig:sigma_xx}, the E-E contribution of $ \text{Im}[\sigma_{xx}(\omega)] $ quickly drops to zero with increasing $ \omega $. However, the corresponding reactive contribution in the Hall conductivity, $ \mathrm{Re}[\sigma_{xy}(\omega)] $, becomes a constant $ e^2/h $  when $ \omega $ exceeds the value of the bulk gap $ 2M $. We find that the flatness of this plateau as a function of $ \omega $ is not sensitive to the system size, as seen by comparing Figs.~\ref{fig:sigma_xy} (a1)-(a3). Adding the three E-E, B-E and B-B contributions, we conclude that the total Hall conductivity in the direct current limit within our finite-size model is $ e^2/h $, rather than $ e^2/2h $ as would have been expected for massive Dirac electrons in an extended system.
	
	Therefore, in the direct current limit, our system behaves as a quantum anomalous Hall insulator with an (integer) Chern number, \textit{i.e.}, a Chern insulator. This motivates the question whether our system behaves as a Chern insulator in response to an \textit{a.c.} field as well. To address this question, we compare the calculated conductivities of our finite-size system for large $ R $ with the conductivities from the low-energy Dirac model $ H=\bm{d}(\bm{k})\cdot\bm{\sigma} $, defined on an extended system. Here we consider two cases and take the spin vector $ \bm{d}(\bm{k}) $ to be linear in momentum with $ \bm{d}_{1}(\bm{k}) = \left(A k_y, -A k_x, M\right) $, and quadratic in momentum with $ \bm{d}_{2}(\bm{k}) = \left(A k_y, -A k_x, M - B(k_x^2+k_y^2)\right) $, where $ A, B(>0),\text{ and } M $ are band parameters independent of momentum. The winding number spanned by the vector $ \bm{d}(\bm{k}) $ is evaluated by this integral
	\begin{align}\label{eq:winding_number}
		\mathcal{C} &= \dfrac{1}{4\pi}\int\mathrm{d}^2k\ \dfrac{\bm{d}\cdot(\partial_{k_x}\bm{d}\times\partial_{k_y}\bm{d}\ )}{d^3}.
	\end{align}
	Accordingly, the quantized Hall response of the linear model, as applicable for the surface states of three-dimensional topological insulators,  is $ \sigma_{xy} = \mathrm{sgn}(M)e^2/2h $ \cite{Qi_RMP}, and that of the second model, as applicable for Chern insulators, is $ \sigma_{xy} = 0 $ when $ M < 0 $ and $ \sigma_{xy} = e^2/h$ when $ M > 0 $.
	
	For extended systems, momentum is a good quantum number and the dynamic conductivity for the above models can be calculated from the Kubo formula \cite{Allen2006} in the momentum representation as usual
	\begin{align}\label{eq:Kubo_2nd}
		\sigma_{\alpha\beta}(\omega) &= \mathrm{i}\hbar\int\dfrac{\mathrm{d}^2k}{(2\pi)^2}\sum_{mn}\dfrac{f_m-f_n}{\epsilon_m-\epsilon_n}\dfrac{\braket{n|j_\alpha|m}\braket{m|j_\beta|n}}{\epsilon_m-\epsilon_n-(\omega+\mathrm{i}\eta)}\\
		&= \dfrac{\mathrm{i} G_0}{4\pi}\int\dfrac{\mathrm{d}^2k}{d}\left[\dfrac{\tilde{j}_\alpha^{-+}\tilde{j}_\beta^{+-}}{\omega-2d+\mathrm{i}\eta}+\dfrac{\left(\tilde{j}_\alpha^{-+}\tilde{j}_\beta^{+-}\right)^*}{\omega+2d+\mathrm{i}\eta}\right]\notag,
	\end{align}
	where $ \tilde{j}_\alpha=\partial_{k_\alpha}H $. Analytic results of the dynamic conductivity tensor for the linear Dirac model can be obtained and is available  \cite{Tse2011PRB}.
	For the quadratic Dirac model, we compute the dynamic conductivity numerically from Eq.~(\ref{eq:Kubo_2nd}) with parameters $ A = \hbar v \simeq \SI{0.3291}{\electronvolt\cdot\nano\meter}, B = \hbar^2/(\SI{2}{\electronmass}) \simeq \SI{0.0381}{\electronvolt\cdot{\nano\meter}^2} $ and $ M = \SI{0.0500}{\electronvolt} $. Interestingly, our results show that not only electronic wavefunctions and dispersions, but boundary conditions can also change the topological property of a system.
	
\begin{figure*}[htp!]
		\centering
		\includegraphics[width=.8\textwidth]{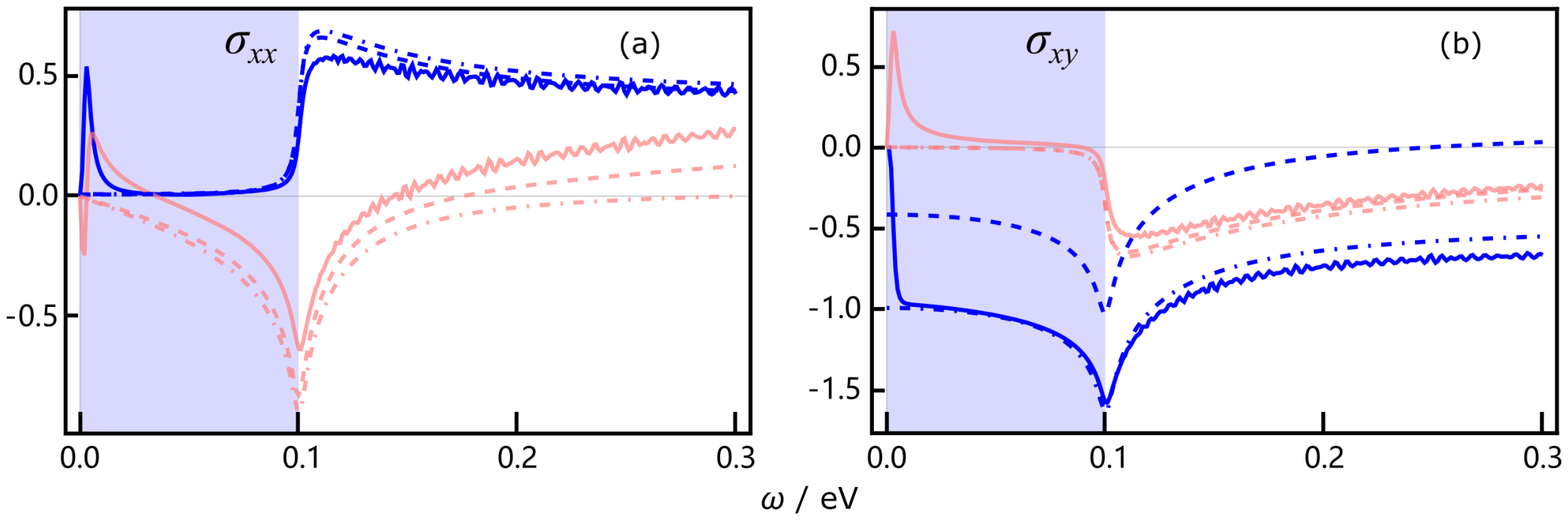}
		\caption{\label{fig:cond_comp} (Color online) Total longitudinal conductivity $ \sigma_{xx} $ [panel (a)] and Hall conductivity $ \sigma_{xy} $ [panel (b)] as a function of frequency, blue (dark) for real part and red (light) for imaginary part. Results from the finite disk (with $ R=\SI{150}{\nano\meter} $) are shown by the thick solid lines, those from the linear Dirac model by the dashed lines and those from the  quadratic Dirac model by dot-dashed lines. Conductivities are expressed in unit of $ G_0 = e^2/h $.}
	\end{figure*}

	Figure \ref{fig:cond_comp} shows the dynamic longitudinal and Hall conductivities for the three cases of finite-sized disk with $ R=\SI{150}{\nano\meter} $, linear and quadratic Dirac models defined on an extended system. For the longitudinal conductivity $ \sigma_{xx} $, we see from panel (a) that the qualitative behavior of all three sets of results resemble each other closely for frequencies beyond the band gap. Within the gap, there is a peak in $ \mathrm{Re}(\sigma_{xx}) $ and a corresponding zero-crossing in $ \mathrm{Im}(\sigma_{xx}) $ near $ \omega = 0 $ in the case of finite-sized disk, which are absent in the extended systems. These features originate from the E-E transition [Fig.~\ref{fig:sigma_xx} (a3)]. For the Hall conductivity $ \sigma_{xy} $, we see from panel (b) that the qualitative behavior the finite-size result matches closely with the extended quadratic Dirac model result, while the extended linear Dirac model result displays a similar trend with increasing frequency but the overall profile is shifted upward. This shift is consistent with the direct current limit of the Hall conductivity for the three cases. Close to $ \omega = 0 $, the finite-size and extended quadratic Dirac models both give a Hall conductivity $ \mathrm{Re}(\sigma_{xy}) $ of $ e^2/h $ whereas the linear Dirac model gives $ e^2/2h $. In addition, a strong peak is apparent near $ \omega = 0 $ in $ \mathrm{Im}(\sigma_{xy}) $ for finite-sized disk due to E-E transition [Fig.~\ref{fig:sigma_xy} (a3)].
	
	\section{Flow Diagram}\label{sec:flow}
	\begin{figure}[htp!]
		\centering
		\includegraphics[width=.45\textwidth]{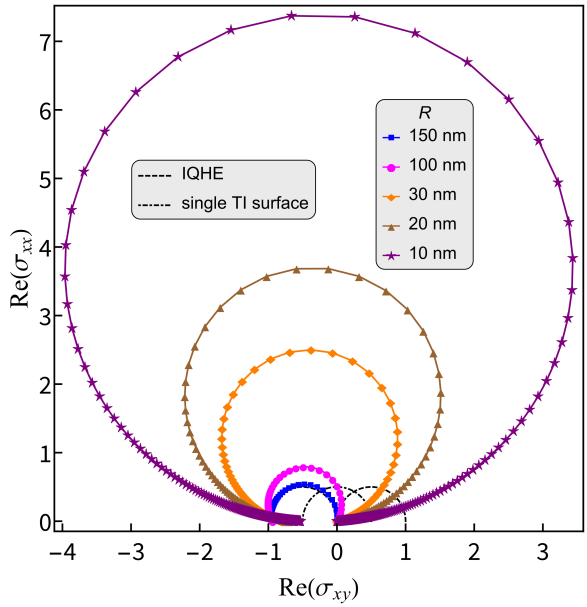}
		\caption{\label{fig:flow_diag} (Color online) Scaling behavior [$ \text{Re}(\sigma_{xx}) $ versus $ \text{Re}(\sigma_{xy}) $] for the E-E contribution. The black dot-dashed line shows the behavior of the topological insulator surface states and the black dashed line the integer quantum Hall state. The various shape-dotted lines show the behavior for the quantum anomalous Hall insulator disk with different radii, which approaches the behavior of the integer quantum Hall state as $ R $ increases. Conductivities are expressed in unit of $ G_0 = e^2/h $.}
	\end{figure}
	Further insights can be obtained by mapping $ \sigma_{xx} $ and $ \sigma_{xy} $ onto a parametric plot using $ \omega $ as the parameter. In the direct current regime, such a $ \sigma_{xx} $-$ \sigma_{xy} $ plot generates a flow diagram and was studied for quantum anomalous Hall insulators theoretically \cite{Nomura2011PRL} and experimentally  \cite{Checkelsky2014NP,Grauer2017PRL} using the system's size, temperature and gate voltage as parameters. The flow diagram of a quantum Hall insulator consists of two semicircles with a unit diameter located at $ (\pm e^2/2h, 0 ) $ \cite{Ruzin,Hilke_1999} [dashed black semicircles in Fig.~\ref{fig:flow_diag}], while the flow diagram of the topological insulator surface states forms a semicircle centered at the origin [dot-dashed black semicircle in Fig.~\ref{fig:flow_diag}]. To study the flow diagram for our finite quantum anomalous Hall insulator disk, we focus on only the E-E contribution of $ \mathrm{Re}(\sigma_{xx}) $ and $ \mathrm{Re}(\sigma_{xy}) $ for frequency $ \omega < 2M $. In this frequency range,  contributions involving bulk states are almost suppressed and thus will not show in the flow diagram. The various shape-dotted lines in Fig.~\ref{fig:flow_diag} show the flow diagram for different values of disk radius from \SIrange{10}{150}{\nano\meter}. The diagram resembles large circular arcs for small $ R $s, and gradually approaches a semicircle centered at $ (-e^2/2h, 0) $ as $ R $ is increased (contrast the purple curve for \SI{10}{\nano\meter} and the blue curve for \SI{150}{\nano\meter}). This is consistent with our finding that the finite disk subjected to the topological infinite mass boundary condition becomes a Chern insulator.
	
	\section{Faraday Effect}\label{sec:F}
	\begin{figure*}[htp!]
		\centering
		\includegraphics[width=.85\textwidth]{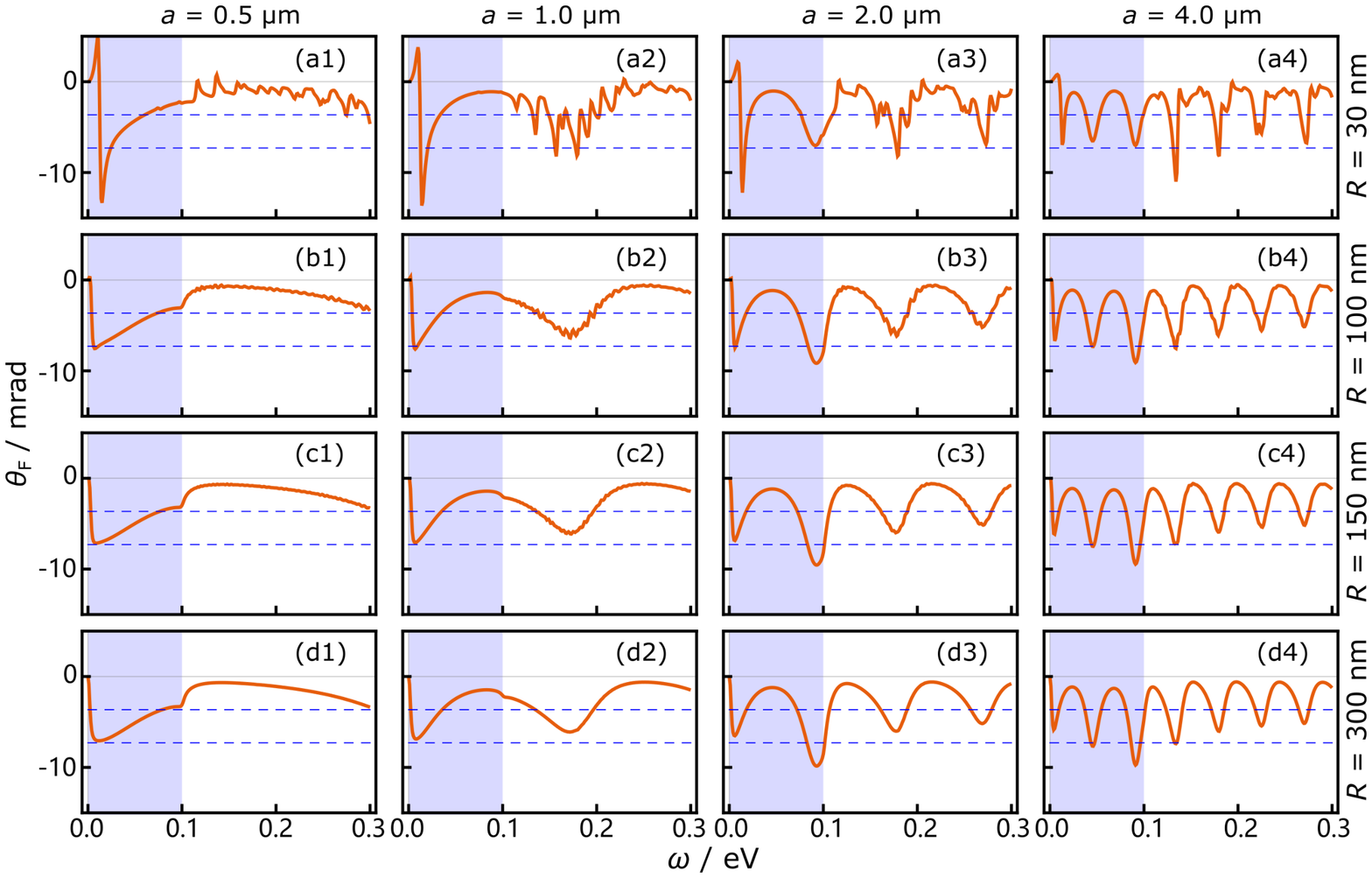}
		\caption{\label{fig:F_rotations} (Color online) Faraday rotation angle $ \theta_\text{F} $ (in milliradian) as a function of $ \omega $ for different values of radius $ R $ and substrate thickness $a$. The light blue (gray) region indicates the extent of the bulk energy gap. The two blue dashed horizontal lines indicate the values of $ -\arctan(\alpha/2)\approx\num{-3.649e-3} $ and $ -\arctan\alpha\approx\num{-7.289e-3} $, with $ \alpha $ being the fine structure constant.}
	\end{figure*}
	
	The dynamic Hall conductivity and its quantized value near the direct current limit can be probed with magneto-optical Faraday effect in topological systems \cite{Tse2010PRL,Tse2010PRB,Tse2011PRB}. We consider a setup consisting of a quantum anomalous Hall insulator nanodisk on top of a dielectric substrate, surrounded by vacuum. For concreteness, we choose silicon \cite{Chang2013AM} as the substrate, which has a dielectric constant $ \varepsilon_{\ce{Si}} = 11.68 $. Using the scattering matrix formalism, the transmitted electric field can be expressed as
	\begin{align}\label{eq:trans}
		{E}^{\mathrm{t}} &= \bar{t}_{\mathrm{B}}\left(\mathbb{I}-\bar{r}_{\mathrm{T}}'\bar{r}_{\mathrm{B}}\right)^{-1}\bar{t}_{\mathrm{T}} \, {E}^{0},
	\end{align}
	where $ E^0 $ is the incident field, $ \bar{r}, \bar{r}' $ and $ \bar{t}, \bar{t}' $ are $ 2\times 2 $ matrices accounting for single-interface reflection and transmission, respectively. The subscript `T' labels the top interface at the quantum anomalous Hall insulator between vacuum and the substrate while `B' labels the bottom interface between the substrate and vacuum. Equation~(\ref{eq:trans}) expresses the transmitted field in the form of geometric series resulting from Fabry-P\'erot--like repeated scattering at the top and bottom interfaces. The incident light is normally illuminated onto the setup and is linearly polarized.
	
	Resolving the transmitted electric field into its positive and negative helicity components $ E_{\pm}^{\mathrm{t}} = E_x^{\mathrm{t}}\pm \mathrm{i} E_y^{\mathrm{t}}$, the Faraday rotation angle is given by
	\begin{align}
		\theta_{\mathrm{F}} &= \dfrac{1}{2}\left(\arg E_+^{\mathrm{t}} - \arg E_-^{\mathrm{t}} \right).
	\end{align}
	
	Our results for the Faraday angle are shown in Fig.~\ref{fig:F_rotations}, computed with different values of disk radius $ R $ and substrate thickness $ a $. $ \theta_{\mathrm{F}} $ shows both finite-size effects due to a finite $ R $ and Fabry-P\'erot resonances \cite{Tse2016} due to the presence of an underlying substrate, which serves as an optical cavity. Let us first focus on the first row (a1)-(a4) with a small radius $ R = \SI{30}{\nano\meter} $. The effect of discrete energy eigenstates in a finite disk is reflected in $ \theta_{\mathrm{F}} $ as fluctuations with frequency [panel (a1)], similar to the behavior seen in the dynamic conductivities. As the substrate thickness is increased [panels (a2)-(a4)], these fluctuations become overwhelmed by the periodic oscillations resulting from the Fabry-P\'erot resonance. For larger radius values across the rows, the fluctuations due to finite disk effect become smoothened and only Fabry-P\'erot oscillations remain. Since the lowest Fabry-P\'erot resonance frequency is $ c/(2a\sqrt{\varepsilon_{\ce{Si}}}) $, more resonances appear for a thicker substrate. When the frequency value is equal to the band gap, we notice a kink in $ \theta_{\mathrm{F}} $ that becomes more apparent for thinner substrates, resulting from the onset and peak features in $ \sigma_{xx} $ and $ \sigma_{xy} $. Another interesting feature is the non-monotonic increase in the envelope of the Fabry-P\'erot oscillations, which become apparent when both $ a $ and $ R $ are large [panels (b4), (c4), and (d4)]. The envelope of oscillations is seen to increase with frequency when the frequency is increased towards the gap and then decreases when the frequency is further increased outside the gap. For large disks, the low-frequency Faraday rotation is closest to the universal value $ -\arctan\alpha\approx\SI{-7.3}{\milli\radian} $ [indicated by the lower dashed line in panels (c1) and (d1)] as predicted for extended systems \cite{Tse2010PRL} when the substrate thickness $ a $ is small, where $ \alpha =  1/137$ is the fine structure constant. Here the value $ \theta_{\mathrm{F}} \simeq -\arctan\alpha $ is consistent with the low-frequency limit of the dynamic Hall conductivity $ \mathrm{Re}(\sigma_{xy}) \simeq e^2/h $. For small radius [panel (a1)], $ \theta_{\mathrm{F}} $ deviates from this universal value due to finite size effects.
	
	\section{Conclusion}\label{sec:disc_conc}
	In conclusion, we have studied the finite-size effects in the dynamical conductivities and magneto-optical Faraday rotation of a quantum anomalous Hall insulator disk. We find that the continuum massive Dirac model subjected to the topological infinite mass boundary condition becomes a Chern insulator with a unit Chern number. In addition, the overall frequency dependence of both the longitudinal and Hall conductivities matches those calculated from an infinitely extended Chern insulator model described by the massive Dirac Hamiltonian with an additional parabolic term. A flow diagram plotting the edge state contribution to $ \mathrm{Re}(\sigma_{xx}) $ versus $ \mathrm{Re}(\sigma_{xy}) $ shows a semicircle in agreement with that expected for a integer quantum Hall insulator. Our numerical results further show that the edge-to-edge transitions constitute the dominant contribution in the Hall conductivity for frequencies not only within but also beyond the bulk band gap. Studies on the magneto-optical Faraday rotation of a small disk show fluctuational features as a function of frequency. These features arise due to optical transitions between  discrete bulk states and is present only for frequencies beyond the band gap. The direct current limit of the Faraday angle shows noticeable deviations with decreasing disk radius from its theoretical value $ \arctan\alpha $ (where $\alpha = 1/137$) based on extended model. In the presence of an underlying substrate, we find an interplay between the Fabry-P\'erot resonances and the finite-size effects due to the size quantization in the Faraday rotation spectrum. Our findings highlight the importance of finite-size effects in optical measurements of dynamic Hall conductivity and Faraday effect in which the laser spot coverage exceeds the sample size. 
	
	\section{Acknowledgment}
	Work at USTC was financially supported by the National Key Research and Development Program (Grant No.s: 2017YFB0405703 and 2016YFA0301700), the National Natural Science Foundation of China (Grant No.s: 11474265 and 11704366), and the China Government Youth 1000-Plan Talent Program. Work at Alabama was supported by startup funds from the University of Alabama and the U.S. Department of Energy, Office of Science, Basic Energy Sciences under Early Career Award {\#}{DE}-{SC}0019326. We are grateful to the supercomputing service of AM-HPC and the Supercomputing Center of USTC for providing the high-performance computing resources.
	
	\appendix
	\section{Infinite mass boundary condition}\label{appsec:IMBC}
	For completeness, here we include the derivation for the infinite mass boundary condition following Ref.~\onlinecite{Berry1987PotRSoLA}. Rather than vanishing wavefunction, this boundary condition requires that the normal component of the current at each point on the boundary vanishes, and for our case this means $ \bm{e}_\rho\cdot\bm{j}(R)=0 $ along the radial direction. For a general two-dimensional massive Dirac model defined on a domain $ \mathcal{D} $ in the real space,
	\begin{align*}
		H= C\mathbb{I} + A\bm{\sigma}\cdot(-\mathrm{i}\nabla) + M\sigma_z,
	\end{align*}
	we consider the total energy $	E=\int_\mathcal{D}\mathrm{d}^2x\ \Psi^\dagger H\Psi$, which can be written as  
	\begin{align}
		E=&\int_\mathcal{D}\mathrm{d}^2x\left[\Psi^\dagger(C\mathbb{I}+M\sigma_z)\Psi\right] \nonumber \\
		&-\mathrm{i} A\int_\mathcal{D}\mathrm{d}^2x\left[\nabla\cdot(\Psi^\dagger\bm{\sigma}\Psi)-\nabla\Psi^\dagger\cdot\bm{\sigma}\Psi\right] \nonumber \\
		=&\left(\int_\mathcal{D}\mathrm{d}^2x\left[\Psi^\dagger(C\mathbb{I}+M\sigma_z)\Psi\right]-\mathrm{i} A\int_\mathcal{D}\mathrm{d}^2x\left[\Psi^\dagger\bm{\sigma}\cdot\nabla\Psi\right]\right)^* \nonumber \\
		&-\mathrm{i} A\oint_{\partial\mathcal{D}}\mathrm{d} l\ \bm{e}_\text{n}\cdot\bm{j}. \label{eq:EJ}
	\end{align}
	In the last equality, Gauss' theorem in two dimensions is used to rewrite the last term into a line integral. Equation~(\ref{eq:EJ}) implies that $ E = E^*-\mathrm{i} A\oint_{\partial\mathcal{D}}\mathrm{d} l\ \bm{e}_\text{n}\cdot\bm{j} $, where $ \bm{j}=\Psi^\dagger\bm{\sigma}\Psi $ is the current operator. Since $ E $ is real, we have $ \bm{e}_\text{n}\cdot\bm{j}=0 $. Next we show how this condition leads to the constraint between the two components of the wavefunction $\Psi=(\Psi_\uparrow,\Psi_\downarrow)^\top $:
	\begin{align*}
		0&=\bm{e}_\rho\cdot\bm{j}(R)=\Psi^\dagger\bm{e}_\rho\cdot\bm{\sigma}\Psi=\Psi^\dagger(\sigma_x\cos\phi+\sigma_y\sin\phi)\Psi\\
		&=
		\begin{pmatrix}
			\Psi_\uparrow^* & \Psi_\downarrow^*
		\end{pmatrix}
		\begin{pmatrix}
			0 & \mathrm{e}^{-\mathrm{i}\phi}\\
			\mathrm{e}^{\mathrm{i}\phi} & 0
		\end{pmatrix}
		\begin{pmatrix}
			\Psi_\uparrow\\
			\Psi_\downarrow
		\end{pmatrix}=2\mathrm{Re}\left\{ \Psi_\downarrow^*\Psi_\uparrow\mathrm{e}^{\mathrm{i}\phi} \right\}.
	\end{align*}
	Therefore, on the boundary, the two components of the wavefunction satisfy $ \Psi_\downarrow/\Psi_\uparrow=\mathrm{i}\alpha\mathrm{e}^{\mathrm{i}\phi} $,
	with $ \alpha=\pm 1 $ for $ \mathrm{sgn}M_\text{in}\ \mathrm{sgn}M_\text{out}=\pm 1 \,$\cite{Berry1987PotRSoLA}, where $ M_\text{in/out} $ represents the mass term for the region of interest (the nanodisk) or the region surrounding it.
	
	\section{Vanishing wavefunction boundary condition}\label{appsec:ZZBC}
	The zigzag boundary condition is used in graphene along a zigzag edge where one of the pseudospin components is taken to vanish. In addition to the fact that there is no microscopic justification for applying the same boundary condition to two-dimensional quantum anomalous Hall insulators, here we also show explicitly that in the massive Dirac model with a finite $ M $ such a boundary condition cannot be applied.
	
	If we try to find edge states by setting one component, say the first one, of the spinor wavefunction in Eq.~(\ref{eq:wave_function}) to vanish on the boundary ($ \tilde{\rho}=1 $) just as what was done in Ref. \cite{Christensen2014PRB}, that means for $ l\neq 0 $, we have $ \epsilon=\pm M $. However, this makes the second component also vanish, so states with $ \epsilon=\pm M $ do not exist.
	
	Then how about the states with $ \epsilon=0 $? To that end, we need to first set $ \epsilon=0 $ and then solve the Schr\"odinger equation which results in a modified Bessel equation as the radial equation. The wavefunction takes the form below [up to a normalization coefficient and $ \kappa=M/(\hbar v)>0 $]
	\begin{align}
		\Phi_{ln}(\rho,\phi)&\sim\mathrm{e}^{\mathrm{i} l\phi}
		\begin{pmatrix*}[r]
			\mathrm{I}_l(\kappa\rho)\\
			-\mathrm{i}\mathrm{e}^{\mathrm{i}\phi}\mathrm{I}_{l+1}(\kappa\rho)
		\end{pmatrix*}.
	\end{align}
	
	For this wavefunction, neither of the component can vanish on the boundary with a finite $ M $.
	
	\section{Derivation of the Kubo Formula in the Real-Space Representation}\label{appsec:Kubo}
	Here we provide a derivation of the real-space Kubo formula used in this paper based on density matrix. Let us start from the Liouville-von Neumann equation $ \mathrm{i}\hbar\partial_t\rho=[H,\rho] $ and consider a system subjected to a time-dependent perturbation $ H=H_0+H'(t)$. In linear response, the density matrix is expanded to first order $ \rho=\rho_0+\delta\rho $, with $\delta\rho $ satisfying $ \mathrm{i}\hbar\partial_t\delta\rho=[H_0,\delta\rho]+[H',\rho_0] $. Assuming a sinusoidal time dependence of $H'(t) \propto \mathrm{e}^{-\mathrm{i}\omega t}$, $\delta\rho $ therefore satisfies $ \hbar\omega\delta\rho = [H_0,\delta\rho]+[H',\rho_0] $. Solving for the matrix element of $\delta\rho $ from above and using the relations $ H_0\ket{m}=\epsilon\ket{m}, \rho_0\ket{m}=f_0(\epsilon)\ket{m} $ gives
	\begin{align}
		\braket{m'|\delta\rho|m} &= \dfrac{f_0(\epsilon')-f_0(\epsilon)}{\epsilon'-\epsilon-\hbar(\omega+\mathrm{i}\eta)}\braket{m'|H'|m},\label{eq:drho_el}
	\end{align}
	where $ f_0 $ is the Fermi-Dirac distribution. Using the summation convention and denoting the elementary charge as $ e=|e|>0 $, the perturbation is $ H' = {e}A_bp_b/m = ev_bE_b/(\mathrm{i}\omega) $. Using the Heisenberg equation of motion, $ \braket{m'|v_a|m} = \braket{m'|\dot{x}_a|m} = {\braket{m'|[x_a,H]|m}}/(\mathrm{i}\hbar) \approx {\braket{m'|x_a|m}(\epsilon-\epsilon')}/(\mathrm{i}\hbar) $. For a two-dimensional system of area $ \mathcal{A} $, the average paramagnetic current density is $ J_a^{\mathrm{p}} = \text{Tr}\{\delta\rho j_a\} $ where $ j_a=(-e)v_a/\mathcal{A} $ is the single-particle current density operator. This gives
	\begin{align}
		J_a^{\mathrm{p}} & =\sum_{mm'}\braket{m'|\delta\rho|m}\braket{m|j_a|m'}    \label{Jpara}  \\ 
		&=-\frac{2\pi}{\hbar} G_0\sum_{mm'}\dfrac{(f_0'-f_0)\Delta\epsilon^2}{\Delta\epsilon-\hbar(\omega+\mathrm{i}\eta)}\dfrac{\braket{m|x_a|m'}\braket{m'|x_b|m}}{\mathcal{A}}A_b, \nonumber 
	\end{align}
	where $ G_0\equiv e^2/h $ is the conductance quantum and $ \Delta\epsilon\equiv\epsilon'-\epsilon $ is the energy difference between two transition states. The paramagnetic current-current correlation function $ \Pi_{ab}(\omega) $, defined through $ J_a^{\mathrm{p}} = \Pi_{ab}(\omega)A_b $, is therefore
	\begin{align}\label{eq:cond_ab}
		\Pi_{ab}(\omega)&=-\frac{2\pi}{\hbar}  G_0\sum_{mm'}\dfrac{(f_0'-f_0)\Delta\epsilon^2}{\Delta\epsilon-\hbar(\omega+\mathrm{i}\eta)}\dfrac{\braket{m|x_a|m'}\braket{m'|x_b|m}}{\mathcal{A}}.
	\end{align}
	
	The conductivity, consisting of both paramagnetic and diamagnetic contributions \cite{Allen2006}, can now be obtained as
	\begin{align}\label{eq:opt_cond}
		\sigma_{ab}(\omega)= & \frac{\Pi_{ab}^{\mathrm{p}}(\omega)-\Pi_{ab}^{\mathrm{p}}(0)}{\mathrm{i}\omega} \nonumber \\
		=& \dfrac{2\pi\mathrm{i}}{\mathcal{A}} G_0\sum_{mm'}\dfrac{(f_0'-f_0)\Delta\omega}{\Delta\omega-\omega-\mathrm{i}\eta}{\braket{m|x_a|m'}\braket{m'|x_b|m}}\nonumber \\
		=& \dfrac{2\pi\mathrm{i}}{\mathcal{A}}G_0\sum_{mm'}(f_0'-f_0)\left(\dfrac{\omega}{\Delta\omega-\omega-\mathrm{i}\eta}+1\right) \nonumber \\
		& \times{\braket{m|x_a|m'}\braket{m'|x_b|m}}\nonumber \\
		=& \dfrac{2\pi\mathrm{i}}{\mathcal{A}}G_0\sum_{mm'}\dfrac{(f_0'-f_0)\omega}{\Delta\omega-\omega-\mathrm{i}\eta}{\braket{m|x_a|m'}\braket{m'|x_b|m}}\nonumber \\
		& +\dfrac{2\pi\mathrm{i}}{\mathcal{A}}G_0\sum_{mm'}(f_0'-f_0)\braket{m|x_a|m'}\braket{m'|x_b|m},
	\end{align}
	where $ \Delta\omega\equiv(\epsilon'-\epsilon)/\hbar $.
	
	The second term vanishes identically because of the commutativity among the components of the coordinate operator,
	\begin{align}\label{eq:opt_cond_ide}
		& \sum_{mm'}(f_0'-f_0)\braket{m|x_a|m'}\braket{m'|x_b|m}\nonumber \\
		= & \sum_{m'} f_0' \braket{m'|x_b x_a|m'} - \sum_m f_0 \braket{m|x_a x_b|m}\nonumber \\
		= & \sum_m f_0 \braket{m|[x_b, x_a]|m} = 0,
	\end{align}
	where in deriving the second line we have used the completeness of states.
	Therefore substituting Eq.~(\ref{eq:opt_cond_ide}) back into Eq.~(\ref{eq:opt_cond}), we obtain the final expression of the conductivity Eq.~(\ref{eq:opt_cond_final}) in the main text with area $ \mathcal{A}=\pi R^2 $ for a disk of radius $ R $,
	\begin{align}\label{eq:opt_cond_final2}
		\tilde\sigma_{ab}(\omega) & =2\mathrm{i}\sum_{mm'}\dfrac{(f_0'-f_0)\omega}{\Delta\omega-\omega-\mathrm{i}\eta}\dfrac{\braket{m|x_a|m'}\braket{m'|x_b|m}}{R^2}.
	\end{align}
	
	\bibliographystyle{apsrev4-2}
	
\end{document}